\documentclass[pdflatex,sn-nature]{sn-jnl}

\usepackage{graphicx}
\usepackage{amsmath,amssymb,amsfonts}
\usepackage{xcolor}
\usepackage{booktabs}
\usepackage{placeins}
\usepackage{float}

\graphicspath{{figures/main/}}

\renewcommand{\Titlefont}{\normalfont\fontsize{16bp}{20bp}\selectfont\bfseries\centering}
\usepackage{etoolbox}
\renewcommand{\Authorfont}{\normalfont\fontsize{11bp}{14bp}\selectfont\centering}
\renewcommand{\addressfont}{\normalfont\fontsize{9.5bp}{12bp}\selectfont\centering}
\makeatletter
\patchcmd{\@maketitle}{\vskip24pt}{\vskip12pt}{}{\PackageError{title-layout}{Title spacing patch failed}{Check the class version.}}
\patchcmd{\@maketitle}{Corresponding author(s). E-mail(s):}{Correspondence:}{}{\PackageError{title-layout}{Correspondence label patch failed}{Check the class version.}}
\patchcmd{\@maketitle}{\vskip21pt}{\vskip0pt}{}{\PackageError{title-layout}{Top spacing patch failed}{Check the class version.}}
\patchcmd{\@maketitle}{\vskip20pt}{\vskip8pt}{}{\PackageError{title-layout}{Title top spacing patch failed}{Check the class version.}}
\patchcmd{\@maketitle}{\vskip20pt}{\vskip12pt}{}{\PackageError{title-layout}{Author spacing patch failed}{Check the class version.}}
\patchcmd{\@maketitle}{\vskip24pt}{\vskip10pt}{}{\PackageError{title-layout}{Abstract spacing patch failed}{Check the class version.}}
\makeatother

\begin{document}

\title{Closed-loop discovery of fluorinated electrolyte additives for high-voltage lithium-metal batteries with expert-guided large language models}

\author[1,2]{\fnm{Haoxiang} \sur{Yin}}\equalcont{These authors contributed equally to this work.}
\author[1]{\fnm{Liwen} \sur{Zhang}}\equalcont{These authors contributed equally to this work.}
\author*[1]{\fnm{Gang} \sur{Wu}}\email{gangwu@scu.edu.cn}
\author*[2]{\fnm{Lixue} \sur{Cheng}}\email{lixuecheng@ust.hk}

\affil[1]{\orgdiv{The Collaborative Innovation Center for Eco-Friendly and Fire-Safety Polymeric Materials (MoE), National Engineering Laboratory of Eco-Friendly Polymeric Materials (Sichuan), State Key Laboratory of Advanced Polymer Materials, College of Chemistry}, \orgname{Sichuan University}, \orgaddress{\city{Chengdu}, \postcode{610064}, \country{China}}}

\affil[2]{\orgdiv{Department of Chemistry}, \orgname{Hong Kong University of Science and Technology}, \orgaddress{\city{Kowloon}, \state{Hong Kong}, \country{China}}}

\abstract{Rising battery demand calls for electrolytes that stabilise lithium-metal anodes and high-voltage cathodes. Additives offer a simple route, yet finding them relies on trial and error. Here we present FLEX, an expert-guided closed loop in which large language models (LLMs) select candidates and explain their choices. Literature-mined fluorinated molecules seeded a space expanded to 24,327 candidates; five LLMs screened it, experts chose molecules to test, and results fed back without retraining. Feedback redirected selection and made explanations more actionable, as rated by experts. Among nine additives, two cyclic anhydrides improved lithium reversibility, protected both electrodes and retained at least 78.9\% capacity after 80 cycles at 4.7~V, versus 11.3\% for the base electrolyte, and over 92\% after 26 pouch-cell cycles. Mechanism analysis linked lithium reversibility more closely to Li$^+$-coordinated reduction potentials than to LUMO energies and distilled two design routes. FLEX shows how a few interpreted experiments can guide additive discovery.}


\renewcommand{\corrauthemail}{\mbox{\href{mailto:gangwu@scu.edu.cn}{\textcolor{black}{gangwu@scu.edu.cn}} (G.W.)};\enspace
\mbox{\href{mailto:lixuecheng@ust.hk}{\textcolor{black}{lixuecheng@ust.hk}} (L.C.)}}

\maketitle

\clearpage
\refstepcounter{section}\label{sec:introduction}

Lithium batteries power electric vehicles and are increasingly used for grid storage \cite{iea2025gevo,iea2024batteries}, at a time when electricity demand is rising, including from data centres for artificial intelligence \cite{iea2025energyai}. The search for higher energy density has renewed interest in lithium-metal batteries \cite{armand2008building,lin2017reviving,weber2019anodefree}. Every component of such a cell matters, but the electrolyte, which contacts both electrodes, largely determines how long the cell lasts: its decomposition at the lithium anode consumes active lithium and leaves electrically isolated lithium, and its oxidation at high voltage degrades the cathode \cite{xu2014electrolytes,fang2019inactive,louli2020diagnosing}. Electrolyte design is therefore central to making lithium-metal batteries practical.

Electrolyte engineering has advanced through concentrated and localised high-concentration formulations \cite{suo2013solventinsalt,yamada2014unusual,chen2018localized} and solvents with tuned solvation \cite{holoubek2021tailoring,cui2024anchoring}. Additives offer the simplest route: a small amount of a well-chosen molecule, often a fluorinated one, can build protective interphases on both electrodes without changing the base electrolyte \cite{zhang2006additives,zheng2017additive,yu2020molecular,zhao2022fluorinated,fan2018nonflammable}. Finding such molecules, however, still relies largely on trial and error, because their behaviour depends on subtle structural features and must ultimately be confirmed in full cells \cite{suo2018fluorinedonating,yu2022rational,choi2025asymmetric,ren2019practical}.

Machine learning has begun to accelerate this search. Data-driven models relate electrolyte composition to lithium reversibility \cite{kim2023datadriven}, physically motivated descriptors guide solvent selection \cite{peng2026rational}, and closed-loop platforms that combine automated experiments with Bayesian optimisation or active learning have identified better electrolytes and charging protocols with fewer experiments \cite{dave2022autonomous,attia2020closedloop,zhu2024differentiable,ma2025active,yang2025dataknowledge,hong2026deep}; learned molecular representations further allow large chemical spaces to be searched \cite{skinnider2021chemical,merchant2023scaling,song2026screening}. These methods are most effective when many measurements can be collected within a defined design space. Early-stage additive discovery is different: each iteration tests only a few molecules, and the most useful information is often a chemical explanation, such as why a molecule helped the lithium anode but failed in the full cell.

Large language models (LLMs) can bring such reasoning into the search. They capture chemical knowledge from the literature \cite{tshitoyan2019unsupervised,wang2023scientific,jablonka2024leveraging}, extract structured data from papers \cite{dagdelen2024structured,polak2024extracting}, plan calculations and experiments as agents \cite{boiko2023autonomous,bran2024chemcrow,ruan2024automatic} and have begun to support catalyst and crystal discovery \cite{song2026digitaltwin,zhang2026adsmind,song2026plausibility}. Because they also explain their choices, their reasoning can be checked against measurements and refined as evidence accumulates. Under expert supervision, such a loop could not only select better candidates but also show which chemical judgements hold and reveal design rules that no single experiment makes apparent. This potential remains to be demonstrated experimentally for electrolyte discovery.

\begin{figure}[H]
\centering
\includegraphics[width=\textwidth]{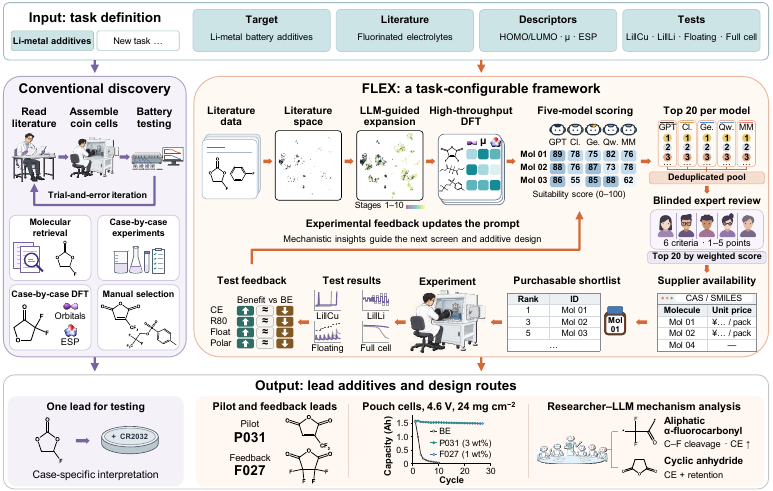}
\caption{\textbf{Overview of the FLEX discovery workflow.} Conventional additive discovery (left) and FLEX (right) start from the same task definition, which specifies the target, literature scope, molecular descriptors and electrochemical tests. In FLEX, literature-derived molecules define a chemical space that an LLM expands, and high-throughput DFT supplies descriptors for five LLMs, which score the library and nominate their top candidates. GPT, Cl., Ge., Qw. and MM denote GPT-5.5, Claude Opus 4.8, Gemini 3.1 Pro (Preview), Qwen3.7-Max and MiniMax M3, respectively. Experts, blind to the nominating models, rank the pooled nominations, purchasable candidates are tested in four electrochemical measurements, and all results, favourable or not, are returned to the models through the prompt for the next screen. The output panel shows the pilot and feedback leads, P031 and F027, in Li$\Vert$NCM811 pouch cells (4.6~V; cathode loading, 24~mg~cm$^{-2}$; 3~wt\% P031 or 1~wt\% F027), and the two design routes derived from researcher--LLM mechanism analysis. Icons and mini-panels in the workflow are schematic; the output panel shows measured data.}
\label{fig:pipeline}
\end{figure}

Here we present FLEX (fluorinated electrolyte exploration), an expert-guided closed-loop workflow for LLM-assisted molecular screening (Fig.~\ref{fig:pipeline}). Five LLMs screen a literature-grounded, PubChem-matched library of fluorinated molecules, experts select candidates for testing, and multi-endpoint measurements return to the models as feedback without retraining. Through pilot and feedback-guided screening and testing, the loop identified two cyclic anhydrides that improved lithium reversibility and sustained 4.7~V cycling, one standing out at the lithium anode and the other at the cathode. A mechanism analysis then tested the models' reasoning against the evidence, moving from overall trends to specific structures. Across the nine tested additives, oxidation descriptors tracked full-cell stability, whereas lithium reversibility followed the reduction potential of the Li$^+$-coordinated additive more closely than the widely used energy of the lowest unoccupied molecular orbital (LUMO). At the molecular level, the analysis distilled two design routes, reductive C--F cleavage in $\alpha$-fluorocarbonyl compounds and a redox signature shared by the cyclic anhydrides, which offer chemical guidance for future additive design.

\section{Results}\label{sec:results}

\subsection{A literature-grounded fluorinated molecular space}

\begin{figure}[H]
\centering
\includegraphics[width=\textwidth]{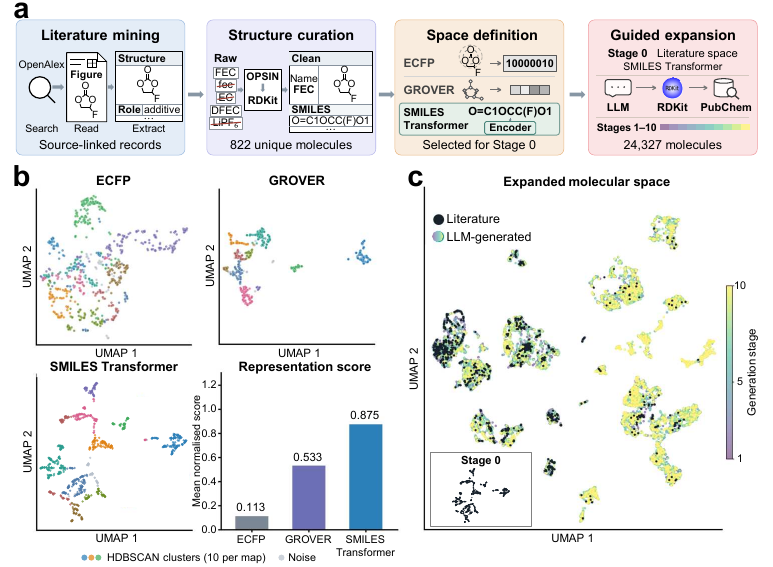}
\caption{\textbf{Definition and staged expansion of the molecular space.} \textbf{a}, Literature mining and structure curation yield 822 unique literature-derived molecules; three molecular representations are compared, and the selected SMILES Transformer space is expanded by an LLM over ten stages, with RDKit checks and PubChem matching. \textbf{b}, The same 822 literature-derived molecules encoded with three representations (ECFP, GROVER and SMILES Transformer) and projected onto two dimensions with UMAP. Each colour is one HDBSCAN cluster; grey points are unclustered. The bar chart scores how well each representation groups related molecules (mean of five min--max-normalised metrics); SMILES Transformer scores highest and was used for expansion. \textbf{c}, Expanded space of 24,327 molecules, projected with all molecules. Literature-derived molecules are shown in black and generated molecules are coloured by expansion stage. The inset (stage 0) shows the literature-derived molecules in the SMILES Transformer projection of \textbf{b}.}
\label{fig:molecular-space}
\end{figure}

Screening needs a candidate space that is grounded in known chemistry yet large enough to contain new additives, so we started from the literature. Mining studies of lithium-metal electrolytes gave 822 unique fluorinated molecules, each linked to the electrolyte role reported for it (Fig.~\ref{fig:molecular-space}a). These molecules cover structural families and applications unevenly (Supplementary Fig.~S1a--f), so we used them as a reference set rather than as the search space itself.

To expand around this reference without losing its chemical relationships, we first compared three molecular representations. SMILES Transformer scored highest, keeping structural families together while spreading the reference molecules across clusters (Fig.~\ref{fig:molecular-space}b; Supplementary Fig.~S2 and Supplementary Table~S1). LLM-guided expansion in this representation, followed by chemical-rule checks and PubChem matching, produced a library of 24,327 fluorinated molecules that fills new regions next to the reported chemistry (Fig.~\ref{fig:molecular-space}c; Supplementary Fig.~S3a,b and Supplementary Table~S2). We then calculated electronic-structure descriptors for this library to support screening.

\newgeometry{textwidth=31pc,textheight=238mm,top=26mm,bindingoffset=6mm,footskip=10mm}

\subsection{Pilot experiments as feedback for LLM screening}

\begin{figure}[H]
\centering
\includegraphics[width=\textwidth]{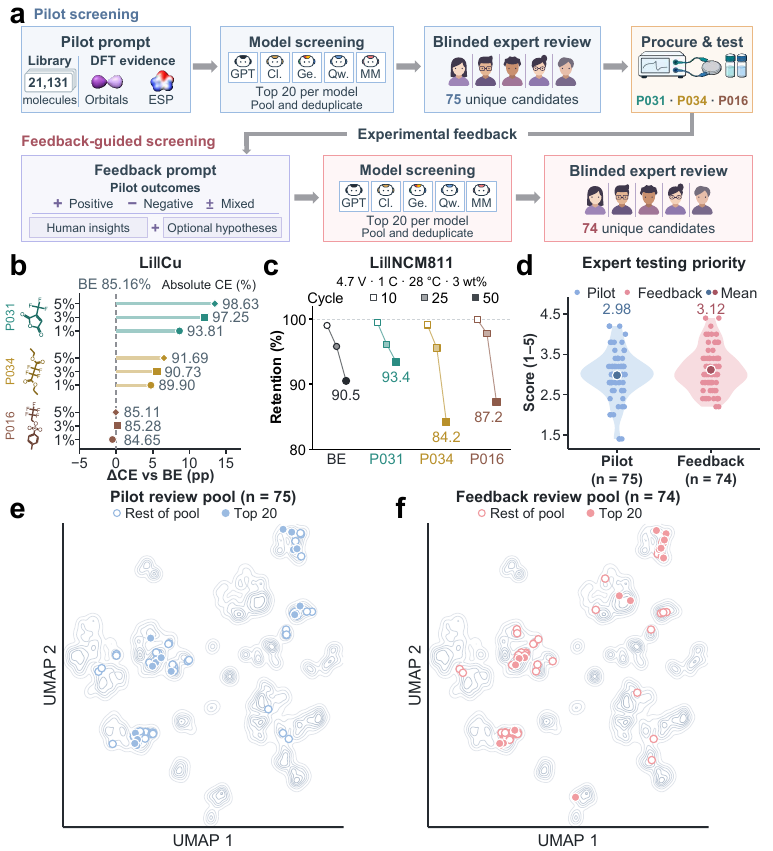}
\caption{\textbf{Experimental feedback changes candidate selection.} \textbf{a}, Pilot and feedback screening. In each stage, five LLMs ranked the library, their top-20 lists were pooled and deduplicated, and five experts, blind to the nominating models, reviewed the pool (75 and 74 candidates). Three pilot additives were procured and tested; their outcomes, together with human insights and an invitation to propose testable hypotheses, formed the Feedback prompt. \textbf{b}, Li$\Vert$Cu CE of the pilot additives at 1, 3 and 5~wt\%, shown as the change relative to BE (85.16\%) in percentage points (pp); labels give absolute CE (\%). \textbf{c}, Capacity retention of Li$\Vert$NCM811 cells with BE or 3~wt\% pilot additive at cycles 10, 25 and 50, relative to cycle 3 (4.7~V, 1C, 28~$^{\circ}$C). \textbf{d}, Expert testing priority of the 75 pilot and 74 feedback candidates (mean of five experts, 1--5 scale); large markers and numbers give group means. \textbf{e,f}, Pilot (\textbf{e}) and feedback (\textbf{f}) review pools in the same UMAP coordinates. Filled markers denote the top 20 candidates by practical score in each pool, and contours show the reference molecular density.}
\label{fig:screening-iteration}
\end{figure}

The pilot screen tested whether the models could nominate useful additives from structures and descriptors alone. Five LLMs ranked the 21,131 library molecules that had complete density functional theory (DFT) descriptors, namely orbital energies, dipole moments and electrostatic potentials (Fig.~\ref{fig:screening-iteration}a; Supplementary Tables~S3 and S4). Experts reviewed the nominations and, taking commercial availability into account, chose three structurally distinct additives: 3-(trifluoromethyl)maleic anhydride (P031), diethyl tetrafluorosuccinate (P034) and 2,2,3,3,3-pentafluoropropyl 4-methylbenzenesulfonate (P016) (Supplementary Table~S5 and Supplementary Fig.~S4).

The three pilot additives behaved differently at the two electrodes, which made them useful feedback. P031 and P034 both raised the lithium plating/stripping Coulombic efficiency (CE) above that of the base electrolyte (BE), whereas P016 showed no consistent gain (Fig.~\ref{fig:screening-iteration}b; Supplementary Fig.~S5). At 3~wt\% and cycle 50, only P031 retained more capacity than BE in full cells (Fig.~\ref{fig:screening-iteration}c; Supplementary Fig.~S6). P034 also lowered Li$\Vert$Li polarisation but drew higher floating currents, a sign of oxidative leakage (Supplementary Figs.~S7 and S8). A single success-or-failure label would have hidden this contrast.

We returned these results to the same five models through a Feedback prompt, without retraining. The prompt combined the three pilot outcomes, five human insights drawn from them and an invitation to state testable hypotheses (Fig.~\ref{fig:screening-iteration}a; Supplementary Note~5). In expert ratings, the full Feedback prompt gave better explanations overall than its ablated versions (Supplementary Fig.~S9). The selections changed as well. Experts rated the new pool slightly higher for testing priority (3.12 versus 2.98; Fig.~\ref{fig:screening-iteration}d), and the Feedback prompt kept the chemical regions of the pilot pool but chose different molecules within them (Fig.~\ref{fig:screening-iteration}e,f; Supplementary Fig.~S10). Six candidates from this pool went on to testing (Supplementary Fig.~S11). Feedback in the prompt, without retraining, thus changed both which molecules the models nominated and how experts rated their explanations.

\subsection{Multi-endpoint testing separates anode and full-cell benefits}

\begin{figure}[H]
\centering
\includegraphics[width=\textwidth]{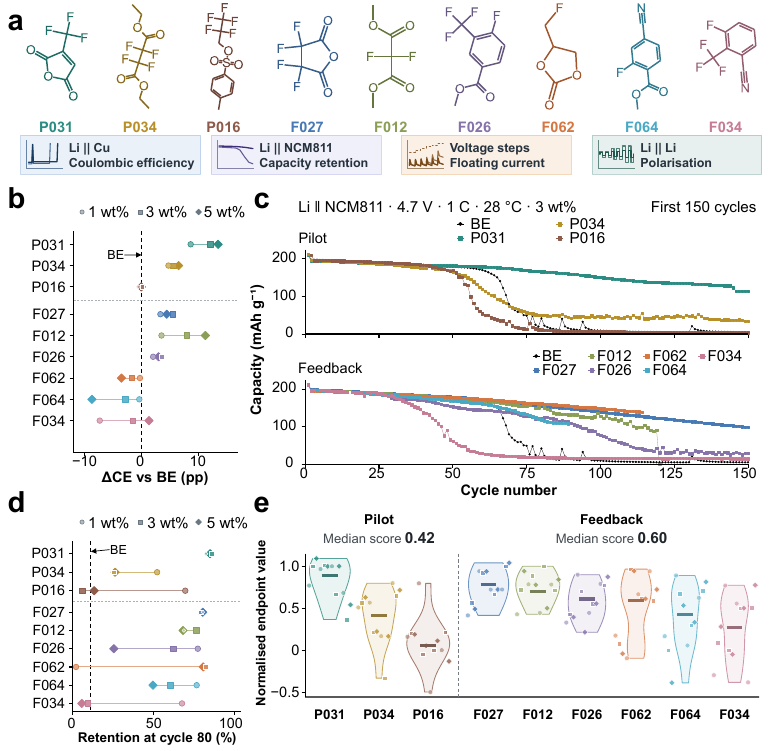}
\caption{\textbf{Complementary endpoints distinguish additive responses.} \textbf{a}, The three pilot (P) and six feedback (F) additives and the four tests (schematic curves): Li$\Vert$Cu CE, Li$\Vert$NCM811 capacity retention, floating current under stepped voltage and Li$\Vert$Li polarisation. \textbf{b}, Li$\Vert$Cu CE change relative to BE (85.16\%) at 1, 3 and 5~wt\%; lines span the three concentrations. \textbf{c}, Li$\Vert$NCM811 cycling with BE or 3~wt\% additive between 3.0 and 4.7~V at 1C and 28~$^{\circ}$C over the first 150 cycles. \textbf{d}, Capacity retention at cycle 80 ($R_{80}$; median of cycles 78--82 relative to cycle 3) at 1, 3 and 5~wt\%; the dashed line marks BE (11.3\%). \textbf{e}, Normalised endpoints of each additive at 1, 3 and 5~wt\%: $\Delta$CE, $R_{80}$, floating current at 4.9~V and Li$\Vert$Li polarisation at 0.4~mA~cm$^{-2}$. Each endpoint is scaled by the range of its concentration medians across the nine additives and BE, with current and polarisation reversed so that higher values are better. Violins span the observed values, dashes mark the equal-weight score of each additive, and the group medians are 0.42 (pilot) and 0.60 (feedback).}
\label{fig:multi-endpoint}
\end{figure}

We then tested the six new candidates and the three pilot additives at 1, 3 and 5~wt\% with four endpoints, each probing a different part of the cell (Fig.~\ref{fig:multi-endpoint}a): CE for lithium reversibility, Li$\Vert$Li polarisation for the voltage cost of lithium cycling, floating current for oxidative leakage and capacity retention after 80 cycles at 4.7~V ($R_{80}$) for full-cell stability. The feedback-guided candidates cycled better: five of six reached a concentration-median $R_{80}$ of at least 60\%, against one of the three pilot additives (Fig.~\ref{fig:multi-endpoint}c,d; Supplementary Figs.~S6 and S12a,b). Two cyclic anhydrides, P031 and tetrafluorosuccinic anhydride (F027), stood out, keeping at least 78.9\% of their capacity at every concentration, whereas the BE cell failed abruptly after about 65 cycles.

The endpoints did not always agree, and their disagreements were informative. P034 improved CE but leaked current under floating and faded in full cells, whereas F062, 4-(fluoromethyl)-1,3-dioxolan-2-one, gave no CE gain yet leaked little and cycled stably at higher loadings (Fig.~\ref{fig:multi-endpoint}b,d; Supplementary Table~S8). Across the nine additives, full-cell retention followed oxidative leakage (Spearman $\rho=-0.83$) more closely than lithium reversibility ($\rho=0.50$), so CE alone would have misjudged both molecules. Low leakage was not sufficient either: at 1~wt\%, F062 lost almost all of its capacity by cycle 80.

The two anhydrides also gave the best balance across all four endpoints: P031 combined the largest CE gains with low leakage, and F027 had the lowest polarisation (30.3 versus 46.5~mV for BE; Supplementary Table~S9). They ranked first and second in the equal-weight score, including when any single endpoint was removed, and the feedback additives scored higher than the pilot additives as a group (median score, 0.60 versus 0.42; Fig.~\ref{fig:multi-endpoint}e; Supplementary Fig.~S12c and Supplementary Tables~S8--S11). Four endpoints thus separated anode benefits from full-cell stability and identified two additives that combined both, which we took forward as the lead electrolytes.

\subsection{The lead electrolytes protect both electrodes}

\begin{figure}[H]
\centering
\includegraphics[width=\textwidth]{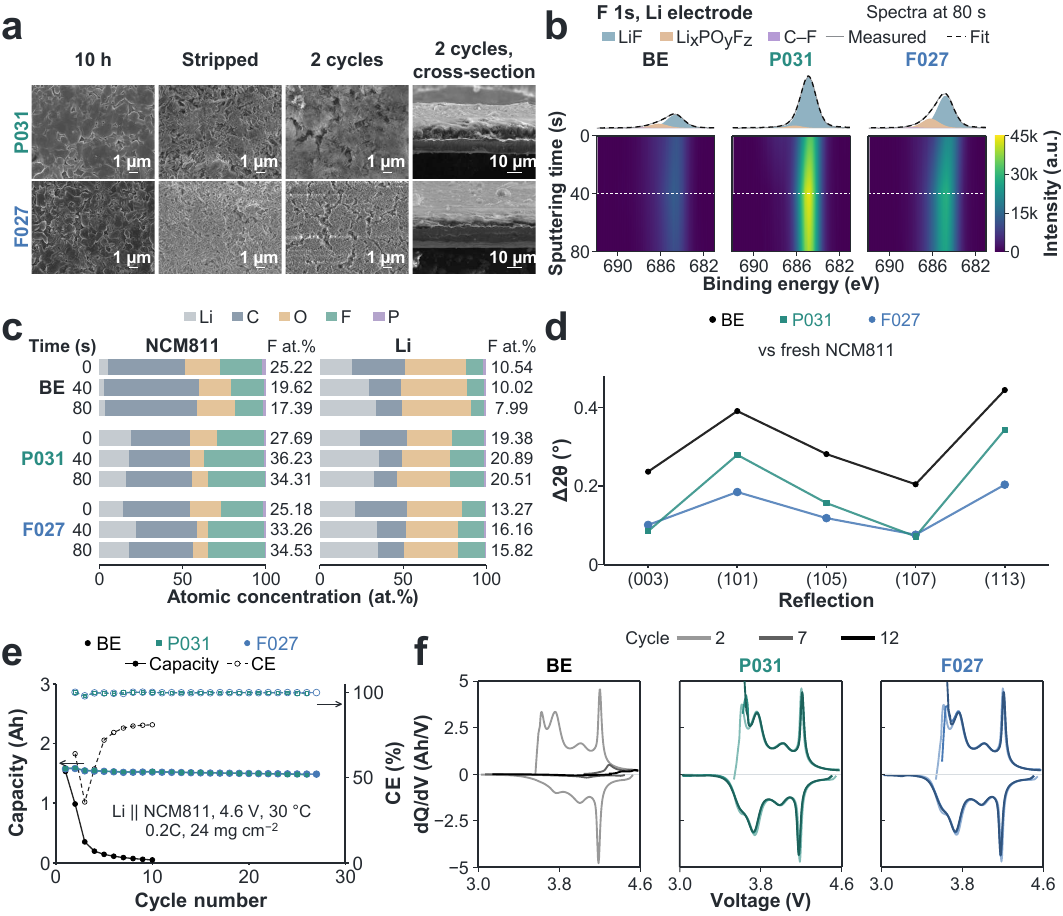}
\caption{\textbf{Interfacial and structural characterisation of the anhydride electrolytes.} BE, 1.0~M LiPF$_6$ in EC/EMC/DMC (1:1:1); P031 and F027, BE with 3~wt\% P031 or 1~wt\% F027. \textbf{a}, SEM images of Li$\Vert$Cu electrodes after 10~h of deposition, after stripping and after two cycles; the right column shows cross-sections after two cycles. Scale bars, 1~$\mu$m (surface) and 10~$\mu$m (cross-section). BE images are shown in Supplementary Figs.~S13 and S14. \textbf{b}, F~1s depth profiles of the Li electrode on a shared intensity scale, with fitted LiF, $\mathrm{Li}_x\mathrm{PO}_y\mathrm{F}_z$ and C--F components shown at 80~s. Intensities between the measured sputtering times (0, 40 and 80~s) are linearly interpolated, and dashed lines mark the measured 40~s depth. \textbf{c}, XPS atomic concentrations of the NCM811 and Li electrodes; numbers give F at.\%. \textbf{d}, NCM811 reflection shifts relative to fresh NCM811, $\Delta2\theta=2\theta(\mathrm{cycled})-2\theta(\mathrm{fresh})$. Electrodes in \textbf{b--d} were recovered from Li$\Vert$NCM811 cells after 20 cycles. \textbf{e}, Discharge capacity (filled symbols) and CE (open symbols, from cycle 2) of Li$\Vert$NCM811 pouch cells (nominal capacity, 1.5~Ah; cathode loading, 24~mg~cm$^{-2}$) cycled between 3.0 and 4.6~V at 0.2C and 30~$^{\circ}$C; BE is shown for the first 10 cycles. \textbf{f}, Differential capacity curves of the same pouch cells at cycles 2, 7 and 12. The sharp peaks at the start of charge extend beyond the plotted range.}
\label{fig:lead-electrolyte}
\end{figure}

We next examined the two lead electrolytes at the electrodes, using each at the concentration that gave its highest R80 (3~wt\% P031 and 1~wt\% F027). The SEM images showed more continuous lithium deposits and less visible residue after stripping with both additives than with BE (Fig.~\ref{fig:lead-electrolyte}a; Supplementary Figs.~S13--S16). Their interphases, however, differed. On cycled lithium, P031 formed the most fluorine-rich interphase, with about 21~at.\% F after sputtering compared with 16~at.\% for F027 and 8--10~at.\% for BE, and its F~1s signal was dominated by LiF, consistent with sacrificial reduction of P031 (Fig.~\ref{fig:lead-electrolyte}b,c; Supplementary Fig.~S17).

On the cathode, F027 gave the smallest mean shift across five NCM811 reflections ($+0.14^\circ$, versus $+0.19^\circ$ for P031 and $+0.31^\circ$ for BE; Fig.~\ref{fig:lead-electrolyte}d). EDS showed fluorine spread evenly over the F027 cathode, and after sputtering its surface remained LiF-rich with the weakest lattice M--O component, consistent with a uniform, LiF-rich interphase that protects the cathode lattice (Supplementary Figs.~S18--S23). In 4.6~V pouch cells, P031 and F027 retained 92.9\% and 94.5\% of their first-discharge capacity after 26 cycles, whereas BE had fallen to 3.1\% by cycle 10 (Fig.~\ref{fig:lead-electrolyte}e). In d$Q$/d$V$, the BE cell lost its peaks by cycle 7, whereas both additive cells kept the NCM811 phase-transition peaks, including the H2--H3 peak near 4.2~V~\cite{ryu2018capacity}, to cycle 12; only the lowest charge peak disappeared as polarisation raised the charge onset (Fig.~\ref{fig:lead-electrolyte}f). Both lead electrolytes therefore improved lithium deposition and high-voltage cycling, with P031 giving the most fluorine-rich lithium interphase and F027 the smallest cathode diffraction shifts. We compared these responses with the calculated redox properties.

\subsection{Mechanism analysis turns test results into design rules}

\begin{figure}[H]
\centering
\includegraphics[width=\textwidth]{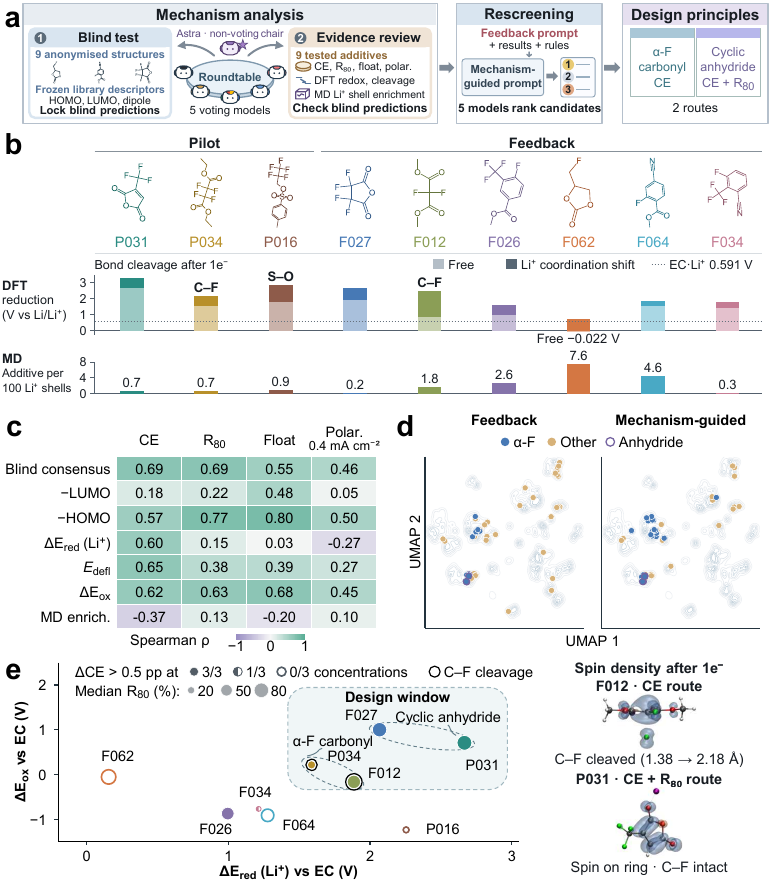}
\caption{\textbf{LLM mechanism analysis links evidence to selection and design.} \textbf{a}, Mechanism analysis and rescreening. Five voting models, with a non-voting chair, first predicted the behaviour of the nine tested additives from anonymised structures and library descriptors (1) and then reviewed the electrochemical, DFT and MD evidence (2). The retained rules were added to the Feedback prompt to form the Mechanism-guided prompt, under which the models rescreened the library. \textbf{b}, DFT reduction potentials of the free additives and their shift on Li$^+$ coordination (V vs Li/Li$^+$); labels mark bonds cleaved in the optimised one-electron-reduced Li$^+$ complexes, and the dotted line marks Li$^+$-coordinated EC (0.591~V). MD bars give the mean number of additive molecules per 100 Li$^+$ first solvation shells at nominal 3~wt\%. \textbf{c}, Spearman correlations across the nine additives between the blind consensus ranking or computed descriptors and the concentration-mean $\Delta$CE and concentration-median $R_{80}$, floating current and polarisation. $E_{\mathrm{defl}}$ is the net reductive-defluorination potential (V vs Li/Li$^+$). Orbital energies, current and polarisation are sign-reversed. \textbf{d}, Untested nominations in the pooled top-20 lists under the Feedback and Mechanism-guided prompts, shown in shared UMAP coordinates; fill marks $\alpha$-fluorocarbonyl structures and outlines mark cyclic anhydrides. \textbf{e}, Li$^+$-coordinated reduction and free-molecule oxidation potentials relative to EC. Fill shows at how many of the three concentrations (1, 3 and 5~wt\%) $\Delta$CE exceeds 0.5 percentage points (3/3, 1/3 or 0/3), marker area scales with median $R_{80}$, and black rings mark C--F cleavage. The shaded window covers $\Delta E_{\mathrm{red}}(\mathrm{Li}^+)\geq1.5$~V and $\Delta E_{\mathrm{ox}}\geq-0.2$~V. Right, spin density of F012 and P031 after one-electron reduction.}
\label{fig:interpretation}
\end{figure}

Selecting better additives was only part of the aim; we also asked whether the models' chemical reasoning holds and whether the results could be turned into design rules. The same five LLMs therefore analysed the nine tested additives in two discussions chaired by a sixth, non-voting model (Fig.~\ref{fig:interpretation}a). They first predicted the behaviour of each additive blind, from its structure and screening descriptors, and then reviewed the measurements together with DFT redox calculations and molecular dynamics (MD) simulations of Li$^+$ solvation. Rules retained in this review were added to the Feedback prompt to form the Mechanism-guided prompt, under which the models rescreened the library.

The blind predictions were largely right about redox chemistry but wrong about solvation. The consensus matched most calculated reduction and oxidation classes, above majority-class baselines, but did no better than the baseline for Li$\Vert$Cu CE (Supplementary Fig.~S24 and Supplementary Table~S17). For solvation, the models expected every additive to be at least as abundant in the first Li$^+$ shell as in the bulk, reasoning from donor groups and polarity. MD showed the opposite: most additives were underrepresented, and none exceeded eight molecules per 100 Li$^+$ shells (Fig.~\ref{fig:interpretation}b). Arguments based on a single molecule did not carry over to the simulated electrolyte.

Comparing descriptors with the measurements showed that reduction and oxidation must be described differently (Fig.~\ref{fig:interpretation}c). For reduction, the environment of the additive mattered. The LUMO energy, widely used to screen for preferential reduction~\cite{halls2010screening}, barely tracked CE gains (Spearman $\rho=0.18$ for $-$LUMO), whereas the reduction potential of the Li$^+$-coordinated additive, $\Delta E_{\mathrm{red}}(\mathrm{Li}^+)$, and the net reductive-defluorination potential, $E_{\mathrm{defl}}$, did ($\rho=0.60$ and 0.65). Coordination raised the calculated reduction potentials by 0.35--1.61~V and reordered the additives, moving F012 from eighth to fourth (Fig.~\ref{fig:interpretation}b); with the same protocol, the correlation with CE rose from 0.35 for the free molecules to 0.60, so the Li$^+$-coordinated potential is the more informative reduction descriptor. The models' blind consensus ranked CE gains about as well ($\rho=0.69$), so their judgement captured more than the free-molecule orbital energy. For oxidation, a simple descriptor sufficed: a deeper highest occupied molecular orbital (HOMO) and a higher oxidation potential, $\Delta E_{\mathrm{ox}}$, accompanied better retention and lower leakage and were associated with all four endpoints, whereas the reduction descriptors related mainly to CE. MD solvation showed no consistent relationship with any endpoint at the low loadings studied.

In optimised one-electron-reduced structures, C--F cleavage appeared only for P034 and dimethyl difluoromalonate (F012), both aliphatic $\alpha$-fluorocarbonyl compounds that improved CE at every concentration (Fig.~\ref{fig:interpretation}b). The models adopted this combination of motif and cleavage as a hypothesis for CE gains (H1). They rejected strong reduction alone as a criterion, because P016 was readily reduced in the calculations without a consistent CE gain, and linked oxidative stability to retention (Supplementary Figs.~S25--S33).

\restoregeometry

Under the Mechanism-guided prompt, the models rescreened the library with the same procedure, and the stated hypothesis spread: the share of $\alpha$-fluorocarbonyl structures among untested nominations rose from 21\% to 48\%, increasing for four of the five models (Fig.~\ref{fig:interpretation}d; Supplementary Table~S12 and Supplementary Fig.~S34). The anhydrides did not spread in the same way. Recorded as two successful molecules rather than as a structural rule, they remained selected, but the number of untested anhydrides in the pooled lists did not change, and most models that kept them still cited low LUMO and deep HOMO energies (Supplementary Fig.~S34g and Supplementary Data~4). Feedback therefore redirected selection most when it was stated as a structural hypothesis.

Together, the evidence review, the correlations and the rescreen point to two design routes (Fig.~\ref{fig:interpretation}e). The CE route pairs an aliphatic $\alpha$-fluorocarbonyl motif with C--F cleavage on the first reduction, as in P034 and F012; after one electron, the C--F bond of F012 breaks. The CE--retention route is represented by the two anhydrides, P031 and F027. They combine the deepest HOMO, lowest LUMO and highest oxidation potentials of the nine additives with strong preferential reduction (Supplementary Fig.~S35), and on reduction P031 keeps its C--F bonds and carries the spin on the anhydride ring (Fig.~\ref{fig:interpretation}e). Only these four additives, spanning both routes, satisfy $\Delta E_{\mathrm{red}}(\mathrm{Li}^+)\geq1.5$~V and $\Delta E_{\mathrm{ox}}\geq-0.2$~V relative to EC. The two sides of the anhydride signature match the two electrodes: P031, the most readily reduced, formed the most fluorine-rich lithium interphase, and F027, the most resistant to oxidation, gave the smallest cathode diffraction shift (Fig.~\ref{fig:lead-electrolyte}). Stated as a rule rather than as two molecules, the signature also points to new candidates: three untested cyclic anhydrides in the library meet its criteria, none of them appeared in the rescreened lists, and DFT confirms that all three lie in the same redox window (Supplementary Fig.~S36b,c).

\FloatBarrier
\section{Discussion}\label{sec:discussion}

Finding electrolyte additives for lithium-metal batteries still relies largely on trial and error. FLEX links LLM screening, expert selection and experiments in a loop that learns from feedback without retraining. Five of the six feedback-guided candidates reached a median R80 of at least 60\%, compared with one of the three pilot additives, and two cyclic anhydrides combined lithium reversibility with stable 4.7~V cycling, retaining more than 92\% of their first-discharge capacity after 26 cycles in 4.6~V pouch cells, against 3\% for BE within 10 cycles. Unlike closed-loop optimisation, which updates predictive models with each measurement~\cite{shields2021bayesian,attia2020closedloop,dave2022autonomous,ma2025active,hong2026deep}, or LLM agents that run automated experiments~\cite{boiko2023autonomous,bran2024chemcrow}, FLEX lets the models propose and explain while experts select and experimentalists test, so a few measurements guide selection through their chemical interpretation.

The loop also made the models' reasoning testable. Their blind consensus matched most calculated redox classes and ranked CE gains more closely than the free-molecule LUMO, but missed every simulated solvation class. Evidence review then clarified which descriptors matter: across the nine additives, CE gains tracked the Li$^+$-coordinated reduction potential more closely than the LUMO, which is often used to screen for preferential reduction~\cite{halls2010screening}, whereas a deeper HOMO accompanied higher full-cell retention. The rescreen also suggested that feedback is taken up more readily as a structural hypothesis than as isolated successes: the $\alpha$-fluorocarbonyl share of untested nominations increased under the Mechanism-guided prompt, whereas untested anhydrides stayed at five.

For electrolyte design, the two routes separate requirements that a single performance label conflates. In the CE route, C--F cleavage in the reduced structures of P034 and F012 accompanied higher CE. In the CE--retention route, the cyclic anhydrides combined strong preferential reduction with oxidative stability, consistent with the fluorine-rich lithium interphase of P031 and the small cathode diffraction shifts of F027. Expressed as computable criteria, this route also identifies three untested anhydrides within the same redox window.

Time and cost limited this work to two experimental iterations and to commercially available molecules. Because the molecular library, code and web platform are openly available, the loop can be extended with further iterations, including tests of the Mechanism-guided prompt nominations and of molecules that require synthesis, and adapted to related discovery problems by changing its task definition. Even at this scale, closing and validating a loop between language models, experts and experiments, and tracing how its selections arose and changed, allowed FLEX to turn a small number of tests into explicit design rules and new directions for additive design.

\section{Methods}\label{sec:methods}

\subsection{Task and chemical scope}\label{task-and-chemical-scope}

The task specified the literature scope, eligible molecular structures, calculated descriptors, screening instructions and experimental endpoints. These inputs and their evaluation criteria defined replaceable modules for literature curation, molecular selection and feedback. The present configuration concerned fluorinated organic additives for lithium-metal electrolytes; model nominations, supplier evidence and expert scores determined the candidates tested.

\subsection{Literature mining and curation}\label{literature-derived-molecular-library}

We searched OpenAlex~\cite{priem2022openalex} for liquid-electrolyte studies
published between 2000 and 2025, combining general electrolyte queries with
searches for fluorinated compounds. Journals were selected for their relevance
to batteries and molecular materials. Language and multimodal models extracted
molecular identities, reported electrolyte roles and supporting references
from article text, tables and figures. Names were resolved to structures using
OPSIN~\cite{lowe2011opsin}, PubChem~\cite{kim2023pubchem} and the NCI Chemical Identifier Resolver.
RDKit was used to standardise structures and remove duplicate identities.
Carbon-containing fluorinated molecules with a reported electrolyte role
formed the literature-derived library used to compare molecular representations
and guide expansion. Search coverage and inclusion criteria are detailed in
Supplementary Note~1.

\subsection{Molecular-space representation}\label{molecular-representation-and-working-space-selection}

Morgan fingerprints (ECFP)~\cite{rogers2010ecfp}, SMILES
Transformer~\cite{honda2019smilestransformer} and GROVER~\cite{rong2020grover}
were compared on the same literature-derived library. Each representation was
reduced to ten dimensions using UMAP~\cite{mcinnes2018umap} and clustered with
HDBSCAN~\cite{mcinnes2017hdbscan}. Separate two-dimensional projections
visualised the space while preserving the cluster assignments.

Representation quality was assessed by agreement with structural families,
local neighbourhood preservation, cluster coverage, separation and balance.
These properties were quantified by family normalised mutual information,
trustworthiness~\cite{venna2001neighborhood}, non-noise coverage,
silhouette~\cite{rousseeuw1987silhouettes} and largest-cluster fraction,
respectively. Metrics were min--max normalised across representations, with
lower largest-cluster fractions favoured, and combined with equal weights.
The highest-scoring representation defined the space for molecular expansion.
Projection and clustering settings are listed in Supplementary Table~S1;
metric definitions are given in Supplementary Note~2.

\subsection{Staged molecular-space expansion}\label{llm-guided-staged-molecular-space-expansion}

GLM-5 received the current molecular space and proposed related structures for inclusion in the library. Proposals targeted four complementary regions: within clusters,
between neighbouring clusters, near sparsely populated boundaries and beyond
the occupied space. Structures were checked with RDKit for chemical validity
and restricted to neutral, single-component, closed-shell fluorinated
molecules. Duplicate identities were removed, and a PubChem identity match
was required for inclusion. Proposals were prioritised by target-region
coverage, structural similarity and novelty relative to the current library.
Accepted structures were mapped into the updated SMILES Transformer space
to guide subsequent stages. The number added at each stage is reported in
Supplementary Table~S2; the expansion procedure is detailed in
Supplementary Note~3.

\subsection{Screening descriptors via DFT}\label{quantum-chemical-calculations}

Quantum-chemical calculations supplied a common set of electronic-structure
descriptors for screening the LLM-proposed structures. Geometry optimisation
and frequency analysis used
B3LYP-D3(BJ)~\cite{becke1993b3,lee1988lyp,grimme2011d3bj}/6-311+G(d,p)
with SMD acetone~\cite{marenich2009smd}. Single-point calculations
used B3LYP-D3(BJ)/def2-TZVP~\cite{weigend2005def2} with the same solvent model.
Orbital energies, dipole moments and molecular-surface electrostatic
potentials were supplied to the screening models. LLM-proposed structures
with an available single-point energy and surface-potential maximum formed
the screening library; literature-derived molecules were excluded from the
model comparison. Further calculation details are provided in
Supplementary Note~4.

\subsection{LLM screening}\label{llm-based-virtual-screening}

Claude Opus 4.8, Gemini 3.1 Pro (Preview), GPT-5.5, MiniMax M3 and
Qwen3.7-Max independently assessed the same screening library under a
common task. Model selection followed external rankings, as
described in Supplementary Note~5. The library was divided into fixed batches of at most 100 molecules.
Each model received molecular
structures, composition, family and cluster information, and DFT descriptors.
Literature roles and proposal rationales were withheld.
The task instructed each model to rank candidates and report suitability
scores, confidence and chemical risks. Nominations were checked for consistent
molecular identities and complete ranks before candidate prioritisation.
Models were queried with the sampling settings listed in Supplementary
Table~S3, using provider defaults where custom values were unsupported.
Prompt conditions are given in Supplementary Table~S4.

\subsection{Candidate prioritisation and expert review}\label{candidate-prioritization-and-blinded-review}

Model nominations were narrowed to an experimental shortlist using supplier
evidence and expert assessment. Candidate ranking combined each model's
suitability score, normalised within-batch rank and confidence. Supplier
evidence was then used to adjust the ranking for commercial availability. For the Mechanism-guided prompt, existing availability annotations were reused and newly nominated candidates were annotated using the same procedure. An additional ranking based on suitability score, local rank and confidence assessed nominations independently of availability (Supplementary Note~6).
The top 20 candidates from each of the five models were pooled and
deduplicated for expert review. Review covered the pilot pool and the Feedback prompt pool.

Five experts assessed molecular structures and calculated properties while
blinded to model identities, scores and support counts. The experts were graduate students, from first-year master’s to second-year doctoral level, with hands-on experimental experience in battery research, including some of the authors; all assessments were blinded to model identity. Six criteria covered
chemical plausibility, suitability as an additive and feasibility of testing.
Ratings on a five-point scale were averaged across experts and combined into
a weighted practical score to select a 20-candidate shortlist. The overall
testing-priority criterion was also summarised separately as the expert
testing-priority score. Experts recorded accept, uncertain or reject decisions.
The scoring rubric and weights are given in Supplementary Table~S5, with
ranking and aggregation rules in Supplementary Note~6. Shortlist rank
and commercial availability guided the selection of candidates for experiments.

\subsection{Experimental feedback and prompt design}

Experimental feedback entered the Feedback prompt, which combined a positive example, the contrasting outcomes, five human insights distilled from the outcomes and a request for chemically grounded, testable hypotheses. Three ablated prompts removed these components in turn: FP-No hypotheses omitted the hypothesis request, FP-Cases only also omitted the insights, and FP-Positive case only kept the positive example alone. The candidate
inputs, batch structure and scoring task were held fixed across conditions
to compare nominations and explanations under different feedback designs.
Candidates nominated under the Feedback prompt entered the feedback-guided expert review.

The Mechanism-guided prompt supplied the four-endpoint experimental records, rules retained by the models and exploratory associations as feedback. The records retained concentrations and test conditions; structural hypotheses and exploratory associations informed assessment rather than imposing eligibility filters. The same five models rescreened the library with the established batches and ranking procedure. The first response meeting the identity and completeness requirements was accepted for each model--batch pair. Each model completed all 212 batches (Supplementary Note~11).

\subsection{Assessment of screening explanations}

Two experts independently rated screening explanations with the same scoring table for evidence grounding, candidate specificity,
mechanistic coherence, risk and uncertainty calibration, experimental
actionability and hypothesis value on a five-point scale. Evaluation materials
concealed model identities and prompt-condition labels. Ratings were averaged
across experts for matched explanations, then across candidate--model sets
with complete ratings under all four conditions. Matching and aggregation
procedures are described in
Supplementary Note~7.

\subsection{Materials and electrolyte preparation}\label{materials}

Battery-grade LiPF$_6$, ethylene carbonate (EC), ethyl methyl carbonate (EMC), dimethyl carbonate (DMC) and LiNi$_{0.8}$Co$_{0.1}$Mn$_{0.1}$O$_2$ (NCM811) cathodes were purchased from Canrd Technology (Dongguan, China). Lithium-metal and copper foils were obtained from China Energy Lithium (Tianjin, China). Polypropylene separators were Celgard 2500 (Celgard, USA).

The additives were 3-(trifluoromethyl)maleic anhydride (P031; CAS 700-27-6; Shanghai Haohong Scientific), diethyl 2,2,3,3-tetrafluorosuccinate (P034; CAS 377-71-9; Shandong Benrite New Chemical Materials) and 2,2,3,3,3-pentafluoropropyl 4-methylbenzenesulfonate (P016; CAS 565-42-4; Shanghai Aladdin Biochemical Technology); dimethyl 2,2-difluoromalonate (F012; CAS 379-95-3; Sichuan Kelun-Biotech Biopharmaceutical), methyl 4-fluoro-3-(trifluoromethyl)benzoate (F026; CAS 176694-36-3; Shanghai Aladdin Biochemical Technology), tetrafluorosuccinic anhydride (F027; CAS 699-30-9; Shanghai Macklin Biochemical Technology), 3-fluoro-2-(trifluoromethyl)benzonitrile (F034; CAS 261951-81-9; Shanghai Aladdin Biochemical Technology), 4-(fluoromethyl)-1,3-dioxolan-2-one (F062; CAS 127213-73-4; Shanghai Macklin Biochemical Technology) and methyl 4-cyano-2-fluorobenzoate (F064; CAS 175596-01-7; Bide Pharmatech). All nine additives, the carbonate solvents and LiPF$_6$ salt were dried and purified before use to meet battery-grade specifications.

NCM811 cathodes were punched into 13~mm discs and vacuum-dried at 80~$^\circ$C for 12~h. The coating contained 94.5~wt\% NCM811 and had a loading of 9.2~mg~cm$^{-2}$. The nominal areal capacity was 1.63~mAh~cm$^{-2}$, based on a specific capacity of 188~mAh~g$^{-1}$. Lithium and copper discs were 15.6 and 19~mm in diameter, respectively. The lithium foil was 450~$\mu$m thick.

BE was 1.0~M LiPF$_6$ in EC/EMC/DMC (1:1:1 by volume). Each additive was introduced separately at 1, 3 or 5~wt\%, defined as additive mass divided by BE mass. The formulations were prepared and kept in an argon-filled glovebox, mixed until homogeneous and used immediately.

\subsection{Electrochemical measurements}\label{electrochemical-measurements}

CR2032 coin cells were assembled with Celgard 2500 separators and 40~$\mu$L of electrolyte, dispensed using a calibrated micropipette, in an argon-filled glovebox with H$_2$O and O$_2$ below 0.1~ppm. Cells were sealed using an MSK-110 crimping machine (Shenzhen Kejing Star Technology) at a load of 0.8~tonnes. After assembly, the cells rested at room temperature for 12~h before testing. Measurements used a Neware battery-testing system at 28~$^\circ$C unless otherwise stated. C-rates were calculated from the cathode active-material mass and nominal specific capacity, with 1C corresponding to a one-hour charge or discharge. Three independent cells were tested for each condition. One representative cell per condition is reported. Where more than one floating-current trace was available for a condition, the cell with the lower median current over the final 10\% of the 4.9~V hold was selected.

Lithium plating/stripping Coulombic efficiency (CE) was measured in Li$\Vert$Cu cells using an adapted Aurbach reservoir protocol with substrate conditioning~\cite{adams2018ce}. All steps used 0.573~mA (0.30~mA~cm$^{-2}$ based on the 15.6-mm Li electrode). Lithium was deposited on Cu for 10~h and stripped to 1.0~V to condition the substrate. After a 2-min rest, a lithium reservoir of $Q_{\mathrm{T}}=5.73$~mAh (3.0~mAh~cm$^{-2}$) was deposited over 10~h, and ten stripping--plating cycles then exchanged $Q_{\mathrm{C}}=1.146$~mAh (0.60~mAh~cm$^{-2}$) with 2-h half-cycles. After a further 2-min rest, residual lithium was stripped to 1.0~V. Average CE was calculated as
\begin{equation}
\mathrm{CE}_{\mathrm{Aurbach}}=\frac{nQ_{\mathrm{C}}+Q_{\mathrm{S}}}{nQ_{\mathrm{C}}+Q_{\mathrm{T}}}\times100\%,
\end{equation}
where $n=10$ and $Q_{\mathrm{S}}$ is the capacity recovered during final stripping. All capacities in the equation were expressed on the same basis. The initial conditioning deposition--stripping cycle was excluded from this calculation.

Li$\Vert$Li symmetric cells were cycled at sequential current densities of 0.1, 0.2, 0.3, 0.4, 0.3 and 0.2~mA~cm$^{-2}$, corresponding to 0.191, 0.382, 0.573, 0.764, 0.573 and 0.382~mA for the 15.6~mm electrodes. The first step comprised ten cycles with 60~min half-cycles. The subsequent five steps each comprised five cycles with half-cycle durations of 30, 15, 7.5, 15 and 30~min, respectively. Each cycle was preceded by a 3~min rest. Extended cycling then used 0.1~mA~cm$^{-2}$ (0.191~mA) with 6~min half-cycles. Polarisation was summarised as the median absolute voltage from 28.5~h up to but excluding 30.0~h, within the 0.4~mA~cm$^{-2}$ step.

Electrochemical floating measurements used Li$\Vert$NCM811 cells charged at a constant current of 0.369~mA to 3.9~V, followed by sequential voltage holds from 4.0 to 4.9~V in 0.1~V increments for 10~h per step. The reported floating current was the median current over the final 10\% of the 4.9~V hold.

Li$\Vert$NCM811 cells were cycled between 3.0 and 4.7~V. After a 5~min rest, cells underwent one formation cycle at 0.2C, followed by another 5~min rest and constant-current cycling at 1C, corresponding to 1.63~mA~cm$^{-2}$. Charge and discharge ended at 4.7 and 3.0~V, respectively, without a constant-voltage stage. Cycle numbering included the formation cycle as cycle 1; the normalisation reference, cycle 3, was therefore the second cycle at 1C. Capacity retention at cycle $n$ was calculated as $100Q_n/Q_3$, where $Q_n$ is the discharge capacity at cycle $n$. Capacity retention at cycle 80 ($R_{80}$) was calculated as 100 times the median discharge capacity over available cycles 78--82 divided by $Q_3$; the record was required to reach cycle 80. Endpoint definitions and acquisition details are given in Supplementary Note~8.

\subsection{Lead-electrolyte pouch cells}

Li$\Vert$NCM811 pouch cells with a nominal capacity of 1.5~Ah used BE, BE with 1~wt\% F027 or BE with 3~wt\% P031. The cells used 20~$\mu$m lithium foil and a 12~$\mu$m polypropylene separator with a 2~$\mu$m ceramic coating. The cathode coating contained 97.4~wt\% NCM811 and had a loading of 24~mg~cm$^{-2}$ and a nominal areal capacity of 4.45~mAh~cm$^{-2}$.

Electrolyte was injected at 0.8~g~Ah$^{-1}$ and the pouch cells were sealed in an argon-filled glovebox with H$_2$O and O$_2$ below 0.1~ppm. An MSK-115A-SS sealing machine (MTI Corporation) was used at a head temperature of 180~$^\circ$C and a vacuum level of $-80$~kPa. The cells rested at room temperature for 48~h after injection and sealing. They were then charged at 0.1C to 4.2~V for formation, degassed and resealed. No electrolyte was replenished during testing.

Cells were cycled between 3.0 and 4.6~V at 0.2C and 30~$^\circ$C. Capacity retention was normalised to the discharge capacity of cycle 1, an activation cycle; cycling CE was reported from cycle 2. Differential capacity ($\mathrm{d}Q/\mathrm{d}V$) curves for cycles 2, 7 and 12 were exported from the Neware battery-testing software.

\subsection{Interfacial characterisation}\label{planned-interfacial-characterization}

Lithium was plated on Cu at 0.573~mA (0.30~mA~cm$^{-2}$) for 5 or 10~h, corresponding to 2.87 or 5.73~mAh (1.5 or 3.0~mAh~cm$^{-2}$), respectively. Additional Cu electrodes were collected after the first strip to 1.0~V and after two complete cycles, each comprising 10~h of plating at 0.573~mA followed by stripping to 1.0~V. Cells were disassembled in an argon-filled glovebox. Electrodes were rinsed with EMC, allowed to dry in the glovebox and transferred under vacuum for scanning electron microscopy (SEM) and energy-dispersive X-ray spectroscopy (EDS) of surfaces and cross-sections.

For post-cycling characterisation, Li$\Vert$NCM811 cells used 40~$\mu$L of BE, BE with 3~wt\% P031 or BE with 1~wt\% F027. Two activation cycles at 0.2C were followed by 20 cycles at 1C between 3.0 and 4.7~V at 28~$^\circ$C. Cycling ended with discharge to 3.0~V. Lithium and NCM811 electrodes were recovered in the glovebox, rinsed with EMC, dried for 3 days and transferred under vacuum. Both electrodes were examined by SEM and X-ray photoelectron spectroscopy (XPS). NCM811 cathodes were also examined by X-ray diffraction (XRD).

SEM used a JSM-7610FPlus field-emission microscope (JEOL) with an ULTIM MAX 40 EDS detector (Oxford Instruments). XRD used a MiniFlex600 diffractometer (Rigaku) with Cu K$\alpha$ radiation at 1.5418~\AA{} and a scan rate of $5^\circ$~min$^{-1}$. XRD patterns were examined over a $2\theta$ range of $5^\circ$--$80^\circ$. Peak positions were determined by local quadratic interpolation over 11 points at a $0.02^\circ$ step, and shifts were calculated relative to fresh NCM811.

XPS used a PHI VersaProbe 4 spectrometer (ULVAC-PHI) with Al K$\alpha$ radiation at 1486.6~eV. Survey scans used 100~eV pass energy and 1~eV steps. High-resolution scans used 50~eV pass energy and 0.1~eV steps. Ar$^+$ sputtering was used only for depth profiles. Atomic concentrations were taken from the instrument quantification. Spectra were referenced to C~1s at 284.8~eV and fitted with a Shirley background and 70:30 Gaussian--Lorentzian profiles, with component widths shared across spectra.
F~1s components were assigned to LiF near 685~eV, fluorophosphates near 686--687~eV and organic C--F near 688~eV~\cite{huang2019supremely,wang2020interface}; O~1s assignments included Li$_2$O near 528~eV on Li~\cite{zhao2022fluorinated} and lattice M--O near 529.5~eV on NCM811~\cite{xu2024proton}. C~1s assignments included C--C/C--H near 284.8~eV, C--O near 286~eV and carbonate species near 289--290~eV~\cite{wang2020interface}, with a PVDF-related contribution on NCM811~\cite{zhao2022fluorinated}; component parameters are provided in the Source Data.

\subsection{Redox and bond cleavage via DFT}

DFT calculations on the tested additives examined reduction, oxidation and
bond cleavage to interpret their electrochemical responses.
Geometry optimisation used B3LYP-D3(BJ)/6-31+G(d,p), followed by
B3LYP-D3(BJ)/6-311+G(d,p) single-point energies with SMD acetone.
Reduction and oxidation potentials were calculated from relaxed-state
Gibbs-energy differences with a common 1.40~V reference offset and were
compared with the corresponding EC reactions. C--F and other bond cleavage after one-electron reduction were identified from the optimised reduced structures. Net defluorination
compared intact $[\mathrm{Li}(A)]^+ + e^-$ with an atom-conserving,
neutral-doublet LiF-contact product, where $A$ denotes the additive.
Gas-phase association energies were evaluated separately. Reaction definitions,
thermal corrections and calculation settings are given in Supplementary Note~9.

\subsection{\texorpdfstring{Li$^+$}{Li+} solvation via MD}\label{atomistic-calculations}

MD simulations evaluated Li$^+$ solvation, additive coordination and first-shell enrichment using GAFF2 for neutral
molecules, Joung--Cheatham Li parameters and CL\&P PF$_6$ parameters.
Solvent and additive charges were fitted to HF/6-31G* electrostatic
potentials using a two-stage restrained electrostatic potential (RESP)
procedure. Neutral charges were unscaled and ion charges were scaled by
0.8. Additives were simulated at nominal 1, 3 and 5~wt\% in
LiPF$_6$/EC/EMC/DMC at 301.15~K. Coordination analysis used
the 2--10~ns production interval and counted each neutral molecule once
in a Li shell if an O or N donor was within 0.30~nm. Additive enrichment
was the frame-averaged ratio of its fraction among organic shell molecules
to its bulk mole fraction among neutral molecules. System compositions,
equilibration, radial distribution functions and analysis settings are
given in Supplementary Note~10.

\subsection{Mechanism analysis by the five LLMs}\label{llm-committee}
In stage A, the same five screening LLMs independently assessed structures, library DFT descriptors and shared test conditions, discussed anonymised peer responses and submitted predictions to a sixth LLM, GPT-6-astra, acting as the non-voting chair. In stage B, the models reviewed DFT and MD evidence together with all four measured endpoints, and the chair consolidated retained rules, exploratory associations and rejected proposals. In stage C, the resulting feedback was incorporated into the Mechanism-guided prompt for full-library screening. Categorical predictions were scored by exact agreement with the reference classes and compared with majority-class baselines; ordinal predictions were compared with measured endpoint rankings (Supplementary Note~11).

\subsection{Candidate redox check via DFT}\label{dft-candidate-check}

We compared the calculated redox properties of untested model nominees and cyclic anhydrides with those of the measured additives. Using the same DFT protocol, we evaluated reduction in the Li$^+$ complex and oxidation of the free molecule relative to EC, and examined the optimised products for bond cleavage. Candidate selection and calculation details are given in Supplementary Note~9.

\subsection{Statistics and reproducibility}\label{statistical-analysis}

All comparisons were descriptive.
Expert testing-priority scores were summarised at the molecule level, with
equal weight assigned to each molecule within a review pool. Molecules
appearing in both pools retained their pool-specific assessments.
Explanation ratings were averaged across the two experts and then across
matched candidate--model sets with complete ratings under all prompt
conditions.

Li$\Vert$Cu CE differences were expressed in percentage points relative to
the common BE value of 85.16\%. At each concentration, $\Delta$CE values
above $+0.5$ percentage points were classified as improved, values below
$-0.5$ as worse, and values within the inclusive interval $[-0.5,+0.5]$
as comparable. The same rule was applied to all additives. A molecule
was classified as improved when all three concentrations improved, mixed
when some improved, and not improved when none improved. Continuous $\Delta$CE values were used in numerical analyses.

For the four-endpoint summary, the median across available concentrations
was calculated for each additive and endpoint. Values were min--max
normalised across the nine additives and BE, with higher CE and $R_{80}$,
and lower floating current and polarisation, assigned higher scores.
The four normalised endpoints were averaged with equal weights. Sensitivity
to omission of each endpoint, use of the 3~wt\% condition alone and use of
the best concentration for each endpoint was evaluated.

Spearman correlations used unrounded values for the nine additives, excluding BE, with average ranks for ties. CE was summarised by the concentration-mean $\Delta$CE; R80, floating current and polarisation used concentration medians. For the evidence--endpoint comparison, floating current and polarisation were sign-reversed so that higher values indicated a favourable endpoint. Library HOMO and LUMO energies were also sign-reversed; calculated redox potentials and MD enrichment retained their physical directions. The models' consensus ordinal rankings were compared with the same endpoint summaries (Supplementary Fig.~S33).

Available same-model score differences between the Feedback prompt and the Mechanism-guided prompt were paired by molecular identity;
model-specific denominators were retained. Motif frequencies were calculated
from each model's fixed top 20 nominations after excluding the nine tested molecular identities and their exact-SMILES aliases, without replacing excluded entries. Pooled lists were deduplicated by candidate identity.
The $\alpha$-fluorocarbonyl and cyclic-anhydride motifs were identified by
SMARTS substructure matching, counting each motif once per molecule and allowing overlap between motifs. Fractions used the parseable untested nominations as their denominator (Supplementary Notes~11 and 12).
Optimised structures and calculation data are provided in Supplementary Data~1--4.

\subsection*{Data Availability}
The data supporting this study are provided in the Supplementary Data, Source Data, the FLEX platform (\url{https://flex.qcatlab.com}) and its repository (\url{https://github.com/AI4QC/FLEX}). Source Data include the interfacial spectra, atomic concentrations, XRD peak positions and cycle-resolved pouch-cell measurements. Raw electrochemical files are available from the corresponding authors on reasonable request.

\subsection*{Code Availability}
The code is available at \url{https://github.com/AI4QC/FLEX} under the MIT licence.

\backmatter

\section*{Acknowledgements}

We thank the HKUST Fok Ying Tung Research Institute and the National Supercomputing Center in Guangzhou Nansha Sub-center for computational resources.

\section*{Declarations}

\subsection*{Competing Interests}
The authors declare no competing interests.

\subsection*{Author Contributions}
H.Y. and L.C. developed the ideas of this work. H.Y. implemented the codes, and L.Z. conducted the experiments. H.Y., L.Z. and L.C. analysed the results and made the graphics and illustrations. L.C. and G.W. supervised the project. All authors contributed to writing the manuscript.

\setlength{\bibsep}{0.60em}

\end{document}